\documentclass[manuscript,nonacm]{acmart}

\usepackage{amsmath,amsfonts}
\usepackage{graphicx}
\usepackage{xcolor}
\usepackage{hyperref}
\usepackage{tabularx}
\usepackage{tabulary}
\usepackage{caption}
\usepackage{subcaption}
\usepackage{adjustbox}
\usepackage{pdflscape}

\newcommand{\xhdr}[1]{\vspace{0.2mm}\noindent{{\bf #1.}}}

\begin{document}

\title{Dimensions of Diversity in Human Perceptions of Algorithmic Fairness}

\author{Nina Grgi\'{c}-Hla\v{c}a}
\affiliation{%
  \institution{Max Planck Institute for Software Systems, Max Planck Institute for Research on Collective Goods}
  \country{Germany}}
\email{nghlaca@mpi-sws.org}

\author{Gabriel Lima}
\affiliation{%
  \institution{KAIST, Institute for Basic Science}
  \country{Republic of Korea}}
\email{gabriel.lima@kaist.ac.kr}

\author{Adrian Weller}
\affiliation{%
  \institution{University of Cambridge, Alan Turing Institute}
  \country{United Kingdom}}
\email{adrian.weller@eng.cam.ac.uk}

\author{Elissa M. Redmiles}
\affiliation{%
  \institution{Max Planck Institute for Software Systems}
  \country{Germany}}
\email{eredmiles@gmail.com}

\begin{abstract}
A growing number of oversight boards and regulatory bodies seek to monitor and govern algorithms that make decisions about people's lives. Prior work has explored how people believe algorithmic decisions should be made, but there is little understanding of how individual factors like sociodemographics or direct experience with a decision-making scenario may affect their ethical views. We take a step toward filling this gap by exploring how people's perceptions of one aspect of procedural algorithmic fairness (the fairness of using particular features in an algorithmic decision) relate to their (i) demographics (age, education, gender, race, political views) and (ii) personal experiences with the algorithmic decision-making scenario. We find that political views and personal experience with the algorithmic decision context significantly influence perceptions about the fairness of using different features for bail decision-making. Drawing on our results, we discuss the implications for stakeholder engagement and algorithmic oversight including the need to consider multiple dimensions of diversity in composing oversight and regulatory bodies.
\end{abstract}

\ccsdesc[500]{Human-centered computing}
\ccsdesc[500]{Human-centered computing~Empirical studies in HCI}

\keywords{Algorithmic Fairness, Fairness in Machine Learning, Human-Centered AI, Human Factors in Machine Learning, Human Perceptions of Fairness}

\maketitle

\section{Introduction}\label{sec:intro}

Algorithms are increasingly used to assist humans with making decisions, in contexts ranging from granting bail \cite{propublica_story} to medical diagnostics \cite{esteva2017dermatologist}. The impact of algorithmic decision-support on human lives has sparked interest in issues of \emph{algorithmic fairness} \cite{agan2016ban, propublica_story, barocas_2016, flores_propublica_reply}. Taking a computational approach, the algorithmic fairness community has proposed various notions of fairness and mechanisms to achieve them \cite{berk2018fairness, corbett2018measure, dwork2012fairness, grgic2018beyond, hardt2016equality, joseph2016fairness, liu2017calibrated, speicher2018unified, zafar_dmt, zafar_fairness}; yet, it has been shown that some of these notions are mutually incompatible \cite{dimpact_fpr, goel_cost_fairness, friedler_impossibility, kleinberg_itcs17} or misaligned with people's perceptions of fairness~\cite{srivastava2019mathematical,saxena2018fairness}.
As it is typically not possible simply to enforce a computational constraint to ensure fairness, there have been increasing calls for algorithmic oversight that addresses the multifaceted ethical, legal, and policy questions involved in using algorithms to help make decisions, from the viewpoint of a broad range of stakeholders.

To help navigate this complex space of ethical and moral issues in AI, the European Commission has formed the ``High-level expert group on AI'' \cite{eu_ai_hleg}, while some corporations have sought to implement oversight boards \cite{Establish:online, Googles97:online, rainie2017theme}. These boards and working groups are composed of research and industry professionals, legal experts, journalists, and human rights activists, amongst others \cite{eu_ai_hleg, oversight_online}. Recent research advocates for further democratizing algorithmic oversight by including not only experts but also those affected by the tools, including the broader public, in such discussions \cite{zimmerman2020technology, rahwan2018society, lee2020human}. Yet, the selection of individuals to staff such boards and working groups has been controversial, in part because of concerns around biases of those holding positions on the boards~\cite{Googles97:online}. With the growing emphasis on ensuring equity, diversity, and inclusion in research \cite{apa2021edi}, academia \cite{cambridge2021edi}, industry \cite{google2021edi}, and beyond, we explore how diversity in experiences and sociodemographics may be relevant to the design, oversight and governance of algorithms utilized in societally consequential domains. 

While much prior work emphasizes the importance of diverse opinions in the discourse about algorithmic fairness \cite{fazelpour2022diversity, jobin2019global, west2019discriminating}, little research has examined \textit{which} dimensions of diversity are most critical. Social psychology research suggests that demographic factors \cite{graham2013moral, thompson1992egocentric, gelfand2002culture} affect people's moral judgements. In this line, a few studies have investigated how people's demographics may introduce biases into their perceptions of algorithmic fairness~\cite{araujo2020ai, pierson2017gender, wang2020factors, albach2021role}. Motivated by the call for inclusion of those affected by algorithmic decisions to oversight and regulatory bodies, we extend previous research by considering a broader range of demographic factors, as well as individuals' personal experiences, which have also been found to influence their moral views \cite{alesina2011preferences, margalit2019political, cassar2017matter, giuliano2014growing, roth2018experienced, gualtieri2018natural}.

We study this question in the context of one aspect of procedural fairness---the fairness of using particular features in an algorithm---in a societally consequential scenario: algorithm-assisted bail decision-making. We run a human-subject study ($n=329$) to evaluate the differences in people's fairness judgements based on (i) demographics (age, education, gender, race, political views) and (ii) personal experience with the algorithmic task being evaluated. 

We find that people's political views are associated with their beliefs about fairness. Although respondents across the political spectrum rank algorithmic features---from most to least fair to use---consistently, left-leaning respondents generally consider using an algorithm for bail decisions less fair than their right-leaning counterparts, regardless of the features used.
Additionally, we find that people who have had personal experiences that are closely related to the decision-making setting judge the fairness of algorithms using certain features differently than those who did not have such experiences. Namely, the experience of having attended a bail hearing is negatively correlated with the perceived fairness of using information about defendants' juvenile criminal history for making bail decisions.

From this analysis, we provide insight into important dimensions of diversity amongst participants in discussions on algorithmic ethics. These findings offer implications not only for the composition of algorithmic oversight boards and regulatory bodies, but for identifying stakeholders to engage in the design and development of algorithms, composing workshop panels, and evaluating the representativeness of the views presented in conversations about algorithmic fairness more broadly.
\section{Related Work}\label{sec:related_work}

Perceptions of fairness have been extensively studied in social psychology \cite{bazerman1995perceptions, colquitt2001dimensionality, colquitt2001justice, greenberg1993social, kahneman1986fairness, yaari1984dividing}. This field of inquiry has been recently extended to encompass \emph{algorithmic} fairness, as reviewed by \citet{starke2021fairness}. Perceptions of algorithmic fairness have been studied in domains such as targeted advertising \cite{plane2017exploring}, lending \cite{saxena2018fairness}, donation allocation \cite{lee2018webuildai}, hiring decisions and work evaluation \cite{lee2018understanding, langer2019highly}, and bail decision-making \cite{grgic2018human, grgic2016case, grgic2018beyond, srivastava2019mathematical, harrison2020empirical}. Most studies have focused on the U.S. population, with few recruiting respondents from other countries, such as Germany \cite{langer2019highly, kieslich2021ai} and the Netherlands \cite{araujo2020ai}.

Many of these studies have found that \emph{people do not reach consensus} in their moral judgments about algorithmic fairness \cite{grgic2018human, albach2021role, starke2021fairness}. Prior work studied the possible causes of this variance: properties of the decision context, the algorithmic decision-aid, and people's individual characteristics.

Fairness judgments have been found to vary with the \emph{decision context}. When comparing the perceived fairness of human and algorithmic decisions, \citet{lee2018understanding} found a preference for human decisions in tasks perceived as requiring human skills, such as work evaluation, but no difference in ``mechanical tasks'', such as work assignment. \citet{nagtegaal2021impact} found that people perceive human decisions as more fair in high-complexity tasks, while favoring algorithmic ones in low-complexity tasks. On the other hand, \citet{araujo2020ai} found a preference for algorithmic decisions in high-stakes health and justice decisions. 

A different line of research studied how fairness judgments vary with respect to the properties of the \emph{decision-support algorithm}. \citet{shin2021effects} found a positive correlation between explainability and perceived fairness, and \citet{binns2018s} found that this effect depends on the type of explanation provided. Other work focused on the perceived fairness of using specific features for making algorithmic predictions \cite{grgic2016case, grgic2018beyond,grgic2018human}. A feature's perceived fairness is positively correlated with its impact on predictive accuracy, perceived relevance, volitionality, and causal relationship with outcomes. On the other hand, features that increase demographic disparity in predictions or are deemed privacy sensitive are perceived as less fair to be used. We build upon this line of work, exploring whether the variance in people's judgments about the fairness of using features can be explained not only by the properties of features, but by people's \emph{individual characteristics}.

\xhdr{Individual Characteristics and Algorithmic Fairness}
\citet{araujo2020ai} studied the association between individual characteristics and the perceived fairness of using algorithms in the health, media, and justice domains. They did not find demographics (age, gender and education) to be significantly associated with fairness judgments. However, the respondents' domain-specific knowledge, beliefs about equality, and online self-efficacy were positively correlated with fairness judgments, while privacy concerns exhibited a negative correlation. \citet{wang2020factors} studied the fairness perceptions of crowdworkers in the context of an algorithm designed to award Master qualifications on MTurk. They found evidence of egocentric effects, with people perceiving algorithms that assign them negative outcomes as less fair. This effect was stronger for women and less educated participants. However, they did not observe a main effect of demographics (gender, education, age) on fairness judgments but only this interaction effect.

In the present study, we focus on the association between people's individual characteristics and their perceptions of \emph{procedural} fairness, namely the perceived fairness of using different features for making algorithmic decisions. Most closely related to our work, \citet{pierson2017gender} explored the impact of gender on perceptions of algorithmic fairness, finding women to be less likely than men to favor including gender as a feature in a system designed to recommend courses to students. \citet{grgic2018beyond}, who considered the same decision context and set of features as we do, descriptively commented on the differences between the extremes on the political spectrum. They found that very liberal respondents perceive the use of features for making bail decisions as less fair than very conservative ones. 

\citet{albach2021role} built upon the work of \citet{grgic2018human}, expanding their study of the impact of features' properties on fairness judgments to six distinct decision contexts: bail, child protective services, hospital resources, insurance rates, loans, and unemployment aid. They found that people's judgments were predominantly consistent across the six decision-making domains, with some domain-specific demographic differences. They examined the relationship between three demographic features: gender, race, and educational attainment on these fairness ratings. Across all six scenarios and features, they found few demographic effects. Of the 4,536 potential relationships they evaluated, only 21 showed a significant correlation with a demographic: race and education. Namely, POC and higher-educated participants rated the fairness of using a few features higher than other respondents. The authors call for further work in this area, given the few relationships they observed and the limited set of factors they analyzed.

In our work, we answer this call. We explore a broader set of individual characteristics, including both a wider set of demographics as well as prior personal experiences, inspired by work in social psychology reviewed below.

\xhdr{Sociodemographics and Moral Judgements}
Past research on Moral Foundations Theory \cite{graham2013moral,graham2011mapping} has found that sociodemographic features such as gender and political views correlate with people's moral views. Studies indicate that women express more concern about fairness-related moral issues than men \cite{efferson2017influence,graham2011mapping}, and that liberals express more concern about such issues than conservatives \cite{graham2013moral,graham2009liberals}. We thus hypothesize similar patterns in our study, looking to see if women and liberals rate certain features as less fair to be used in algorithmic decision-support.

To form hypotheses about political leaning, we can further refer to research on individualist and structuralist beliefs. Compared to liberals, conservatives are more likely to attribute poverty and criminal behavior to individualistic factors---which are under a person's control---than to societal causes, which are beyond one's control \cite{hunt2004race, zucker1993conservatism, carroll1987sentencing}. Much prior work has discussed the relationship between the degree of control over outcomes and fairness. Luck egalitarianism argues that people's outcomes should be determined based on their choices, and not on brute luck \cite{anderson1999point, knight2013luck}. Research on the deservingness heuristic has found that people favor allocating social welfare to those they perceive as being unlucky rather than lazy \cite{petersen2011deservingness}. Finally, \citet{grgic2018human} found that the perceived volitionality of a feature is positively correlated with its perceived fairness of use in algorithmic decision-support. Hence, as stated above, we hypothesize that conservatism may be positively correlated with the perceived fairness of features. 

Additionally, some research found correlations between other sociodemographic factors and perceptions about fairness. African-Americans are more likely to perceive the criminal justice system as unfair \cite{hurwitz2005explaining}. Younger adults are more likely to believe that computer programs can be free from bias \cite{smith2018public}. Educational attainment was found to be positively correlated with fairness as considered by Moral Foundations Theory \cite{van2014moral}. Hence, we also conduct an exploratory analysis of the association between perceptions of fairness and the respondents' race, age, and education.

Research on egocentric interpretations of fairness \cite{thompson1992egocentric} suggests that egocentricity may effect people's fairness judgments, especially in individualistic societies \cite{gelfand2002culture} such as the U.S. Accordingly, we hypothesize that people's perceptions of the fairness of using specific features, such as age, race or gender, may vary egocentrically based on the respondents' age, race and gender, respectively. Namely, disadvantaged groups (POC, younger individuals, and women\footnote{In the context of the COMPAS tool, POC, younger individuals and women can be considered disadvantaged groups, since those individual attributes are significantly correlated with receiving higher risk estimates. Additionally, the COMPAS tool utilized in Broward County, Florida was found to overestimate the criminal recidivism risk of women and POC\cite{larson2016}.}) may perceive the use of the corresponding features in algorithmic decision-support as less fair. Alternatively, instead of this ego-justifying and group-justifying behavior, participants may engage in system-justifying behavior \cite{jost2019quarter}, studied by System Justification Theory. Disadvantaged individuals and groups are found to sometimes exhibit outgroup favouritism and perpetuate negative stereotypes about themselves \cite{jost1994role, jost2019quarter}, while justifying the status quo which puts them in a disadvantaged position. Hence, we alternatively hypothesize that perceptions of fairness may vary in a manner opposite to the egocentric direction, exhibiting outgroup favoritism.

\xhdr{Personal Experiences and Fairness Preferences}
Prior work found that people's perceptions of fairness correlate with their past experiences. Namely, negative and traumatic past experiences at both the \emph{individual} and \emph{societal} levels are associated with greater support for fairness interventions.

\citet{alesina2011preferences} found that factors related to a person's past experiences, such as experiencing unemployment and personal traumas, are positively correlated with their support for wealth redistribution. Similarly, \citet{margalit2019political} found evidence that economic shocks, such as job loss or a sharp drop in income, tend to increase support for more expansive social policies. \citet{cassar2017matter} found that participants who experienced an economic failure in a lab experiment were more likely to favor redistribution, even in the absence of personal monetary stakes. Collectively, these studies show that \emph{individual}-level experiences impact people's perceptions of fairness.

Other studies explored the effects of society-level experiences. \citet{giuliano2014growing} found that an individual's experience of an economic recession while growing up is positively correlated with support for government-led wealth redistribution. In contrast, \citet{roth2018experienced} showed that people who grew up in times of higher economic inequality are less likely to consider the current real-world income distribution unfair and support wealth redistribution. \citet{gualtieri2018natural} found that experiencing natural disasters also affects fairness preferences---the intensity of the shakes that people felt during the 2009 l'Aquila earthquake is positively correlated with their support for redistribution.

In our work, we leverage research on the effects of personal experiences and egocentric effects and consider a specific subset of individual characteristics that may exhibit both effects: experiences closely related to the decision-making scenario. We additionally conduct an exploratory study, to explore if respondents who have personal experience with the decision-making task make fairness judgments differently than those who do not.
\section{Methodology} \label{sec:methodology}
We use a quantitative survey ($n=329$) to assess the relationship between respondents' individual factors and their perceptions of procedural fairness in an already well-studied algorithmic support context: bail decision-making. Below, we list our hypotheses, and describe our survey instrument, sampling procedures, analyses, and the limitations of our work. The procedures we describe were approved by our institution's IRB board.

\subsection{Hypotheses}

We leverage prior work reviewed in Section \ref{sec:related_work} to form the following hypotheses:

\noindent \textbf{Hypothesis 1---Political Leaning:} On a scale from very liberal to very conservative, people's political leaning is positively correlated with their fairness ratings.

\noindent \textbf{Hypothesis 2---Gender:} Women rate the use of features as less fair than men.

\noindent \textbf{Hypothesis 3a)---Ego- and group-justifying behavior:} Disadvantaged groups (women, POC, and younger individuals) rate the use of corresponding features (gender, race, and age respectively) as \emph{less fair} than those who are not members of the disadvantaged group.

\noindent \textbf{Hypothesis 3b)---System-justifying behavior:} Disadvantaged groups (women, POC, and younger individuals) rate the use of corresponding features (gender, race, and age respectively) as \emph{more fair} than those who are not members of the disadvantaged group.

We additionally conduct an exploratory study about the relationship between people's perceptions of algorithmic fairness, and a broader set of sociodemographic factors (age, education, and race) and lived experiences.

\subsection{Survey Instrument} 
Survey respondents were presented with an algorithmic bail decision-making scenario inspired by the COMPAS tool, which assists judicial decisions in several U.S. jurisdictions by estimating defendants' risk of criminal recidivism \cite{propublica_story}. Using this scenario, we queried respondents' perceptions about the fairness of using eight features from the ProPublica dataset~\cite{propublica_story}, which contains information about more than 7000 criminal defendants who were arrested and subsequently subjected to COMPAS screening in Broward County, Florida in 2013 and 2014. The features about which we asked respondents have received considerable attention from previous work \cite{zafar2017fairness, grgic2018beyond, dressel2018accuracy} and capture information about the defendants' (1) \emph{number of prior offenses}, (2) \emph{precise description of the current arrest charge}, (3) \emph{degree of the current arrest charge degree}, (4) \emph{number of juvenile felonies}, (5) \emph{number of juvenile misdemeanors}, (6) \emph{age}, (7) \emph{gender}, and (8) \emph{race}. 

For each of the eight features, respondents were shown a description of the scenario and asked whether they agreed that it was fair to use the feature for bail decisions using a 7-point Likert scale (1 = ``Strongly Disagree,'' 7 = ``Strongly Agree''). The responses to these questions are the study's dependent variables. 

Following survey methodology best practices~\cite{redmiles2017summary}, which suggest re-using previously used and already pre-tested survey questions in future research, we draw the phrasing of the scenario and the fairness perception questions from the pre-validated approach of \citet{grgic2018human}. Figure \ref{fig:question} in the Appendix shows an example vignette for the feature \emph{race}. The eight vignettes were shown in random order to avoid order bias \cite{redmiles2017summary, groves2011survey}. Respondents then answered two attention-check questions, which we used for quality assurance. The attention-check questions were instructed-response items, in which respondents were instructed to select a specific response option in a multiple-choice question. Similar questions are commonly employed for identifying inattentive respondents in online surveys \cite{meade2012identifying}.

Next, respondents answered a series of questions concerning their personal experiences. Respondents were asked whether they had (1) heard or read anything related to COMPAS before taking the survey; if (2) they or (3) their close friends or relatives held a job or have education in a law or crime-related field; and if they had ever (4) attended a bail hearing or (5) served on a jury. The considered personal experiences vary with respect to their closeness to the task at hand, ranging from close experiences with bail decisions (4), and close experiences with the legal system (2 and 5) to superficial familiarity with the decision context (1) or the legal system (3). Finally, we gathered data about respondents' demographics: age, gender, race, education, and political leaning. The experience questions were shown first in random order, followed by demographic questions shown in random order. Both sets of questions were optional, i.e., respondents had the option to opt out of responding to them. These responses comprise the study's independent variables. The exact phrasing of the aforementioned survey questions is listed in Table \ref{tbl:survey_questions} in the Appendix.

Finally, we concluded the survey by asking respondents to share their thoughts and feelings about participating in this study,\footnote{These questions were inspired by the \emph{enjoyment}, \emph{ease of responding} and \emph{intention to respond to a similar future survey} scales introduced by \cite{croteau2010employee}, and used in \cite{huang2015detecting}. For all four questions, we gathered responses using a 5-point Likert scale, from ``Strongly agree'' to ``Strongly disagree'.'} including how interesting they found it, if they would be willing to partake in a similar study in the future, and how difficult they found the questions to understand and respond to.

\begin{table}[t]
\parbox{.45\linewidth}{
\centering
    \begin{tabulary}{\linewidth}{L|R|R}
    \toprule
    {\bf Demographic Attribute} & {\bf Sample} & {\bf Census}\\
    \hline
    <35 years & 53.8\% & 46\%\\
    35-54 years & 35.6\% & 26\%\\
    55+ years &  10.6\% & 28\%\\
    \hline
    Male & 48\% & 49\%\\
    \hline
    Asian & 10.3\% & 6\% \\
    Black & 5.8\% & 12\% \\
    Hispanic & 7.6\% & 18\% \\
    White & 73.6\% & 61\%\\
    Other & 2.7\% & 4\% \\
    \hline
    Liberal / Democrat & 50.8\% & 33\%$\dagger$\\
    Moderate / Independent & 24.3\% & 34\%$\dagger$\\
    Conservative / Republican & 24.9\% & 29\%$\dagger$\\
    \hline
    Bachelor's or above & 60.5\% & 30\%\\
    \bottomrule
    \end{tabulary}
    \caption{Demographics of our survey sample, compared to the 2019 U.S. Census~\cite{census_acs}. Attributes marked with a $\dagger$ were compared to Pew data~\cite{pew_politics} on political leaning from 2016.}
    \label{tbl:demographics}
    
    \begin{tabulary}{\linewidth}{L|R}
    \toprule
    {\bf Personal Experiences} & {\bf Sample} \\
    \hline
    Heard of scenario & 6.7\%\\
    Legal profession -- you & 7.6\%\\
    Legal profession -- friends \& relatives & 23.1\%\\
    Attended bail hearing & 14.6\%\\
    Served on jury & 22.5\%\\
    \bottomrule
    \end{tabulary}
    \caption{Prior personal experiences of our respondents.}
    \label{tbl:experiences}
    }
\hfill
\parbox{.45\linewidth}{
    \centering
\def\sym#1{\ifmmode^{#1}\else\(^{#1}\)\fi}
\begin{tabulary}{\linewidth}{L|R|R}
\toprule
& {\bf Coef.} & {\bf SE} \\
\hline
Age & 0.0494 & (0.0418)\\
Political leaning & \textbf{0.236}\sym{***}& (0.0454)\\
Bachelor's or above & 0.0297 & (0.107)\\
Male & 0.0492 & (0.103)\\
White & -0.0940 & (0.116)\\
\hline
Heard of scenario & -0.0312 & (0.208)\\
Legal profession - you & -0.0907 & (0.206)\\
Legal profession - fr \& rel & -0.0143 & (0.126)\\
Attended bail hearing & -0.256 & (0.148)\\
Served on jury & 0.0432 & (0.127)\\
\hline
Constant & \textbf{3.668}\sym{***}& (0.160)\\
\bottomrule
\end{tabulary}
\caption{Linear mixed model with a random effects term for respondents. Dependent variable: fairness ratings on a 7-point Likert scale with larger values indicating higher perceived fairness. Independent variables (rows): respondents' \textbf{demographics and experiences}. Reference groups for the ordinal variables age and political leaning are ``18-24'' and ``very liberal'' respectively. Number of observations = 2632. Standard errors in parentheses. *** p < .001, ** p < .01, * p < .05.}
\label{tbl:regression_across_feats}}
\end{table}

\subsection{Sampling}
We deployed the survey on the online crowdworking platform Prolific~\cite{palan2018prolific}. Prolific is a platform which offers services explicitly targeted at researchers, including a plethora of fine-grained criteria for pre-screening respondents.

\xhdr{Pre-screening Criteria}
We used several pre-screening criteria to ensure data quality, per best-practice research guidelines~\cite{eyal2021data}. We targeted respondents who self-reported English fluency and had participated in at least ten previous studies on Prolific, with an approval rate above $95\%$. Additionally, we only recruited respondents located in the U.S. to ensure that respondents have a basic understanding of the U.S. legal system. Finally, we used additional pre-screening criteria to gather a more representative sample by recruiting more respondents with under-represented demographics and experiences. Namely, we deployed additional studies targeting right-leaning respondents, who are typically under-represented on online crowdworking platforms \cite{huff2015these}, and men, who are currently under-represented on Prolific \cite{prolific_gender_balance}, as well as people who self-reported to Prolific to have served on a jury, been the victim of a crime, or been to prison. This targeting was not conducted via pre-screening surveys, but instead using Prolific's interface that offers these pre-screening categories, and this sensitive data about participants was not retained nor used in our analysis.

\xhdr{Sample Size Rationale}
To estimate the required sample size, we conducted a power analysis using the software G*Power \cite{faul2007g}. We accounted for the fact that our sample is likely to be imbalanced in terms of many of the demographics and experiences we consider in our study. While we expected a balanced sample with respect to gender, we anticipated fewer respondents who have served on a jury and even fewer who work in a legal profession. Our goal was to attain a sample that would enable us to identify medium-sized effects (Cohen's $d=0.5$) for minorities that constitute at least $15\%$ of the sample, and large-sized effects (Cohen's $d=0.8$) for minorities that constitute at least $5\%$ of the sample. In the context of our correlational study, the effect size captures the magnitude of the difference between the fairness judgments of respondents who self-identify as having different individual characteristics. We focused only on medium- and large-sized effects because small effects may arguably be practically insignificant in our context, regardless of their statistical significance. Based on the two-tailed Wilcoxon-Mann-Whitney test, with standard values $\alpha=0.05$ and power ($1-\beta$) 0.8, the minimum required sample size for identifying medium-sized effects for $\geq 15\%$ minorities is a total of 260 respondents (39 minority group and 221 majority group). For detecting large effects for $\geq 5\%$ minorities, the minimum required sample size is slightly larger: a total of 274 respondents (14 minority group and 260 majority group).

\xhdr{Respondents}
We gathered responses from 363 respondents during December 2021. We sampled respondents at different times of the day and on five days of the week to reduce sampling bias that may occur due to the day in the week or the time of day \cite{casey2017intertemporal}. We removed data from 9 respondents who provided incorrect responses to either of the two attention check questions. Additionally, we discarded all data gathered from 13 respondents who opted not to respond to the demographic or experience-related questions. Finally, we discarded data from 12 respondents who reported their political leaning as ``Other,'' to preserve the ordinal structure of the variable for our analysis. Our final sample consisted of 329 respondents. All respondents were paid 0.8 GBP (approx. \$1.08 USD) for completing the study, which took an average of 5.2 minutes. The average hourly rate was hence approximately \$12.50.

Table \ref{tbl:demographics} shows the demographics of our sample, compared with the 2019 U.S. Census~\cite{census_acs} and 2016 Pew data on political leaning~\cite{pew_politics}. Compared to the U.S. census, our sample is younger, more educated, more liberal leaning, and consists of more white respondents, as is typically the case for samples recruited on online crowdworking platforms \cite{huff2015these, ross2010crowdworkers, paolacci2010running}. Table \ref{tbl:experiences} details the respondents' personal experiences related to the decision-making task. Most respondents had not heard of COMPAS nor had they been involved in bail decision-making or juries.

The overwhelming majority of respondents expressed positive sentiments about participating in the study. 93\% of respondents found the survey interesting, and 98\% stated that they would like to take part in a similar survey in the future. Less than 1\% of respondents found the questions difficult to understand, and 4\% found them difficult to answer.

\subsection{Analysis} \label{subsec:analysis}
We analyzed our data with linear regression models. People's individual characteristics were treated as independent variables, while the 7-point Likert scale fairness ratings were treated as dependent variables. The independent variables are either binary (experiences), ordinal (political leaning: 5-point Likert scale from ``Very liberal'' to ``Very conservative'' coded as numerical values from 0 to 4; age: buckets ``18-24'', ``25-34'', ..., ``85 or older'' coded as numerical values from 0 to 7), or transformed to binary (gender: male vs non-male; race: white vs non-white; education: Bachelor's and above vs below Bachelor's). The specifics of each model are described alongside their results in Section \ref{sec:results}.

\xhdr{Limitations}
In this paper, we study how people's individual characteristics are associated with their fairness judgments. Individual characteristics, such as demographics and life experiences, are not experimental conditions to which respondents were randomly assigned. Hence, we do not make claims about the causal effects of people's individual characteristics on their fairness judgments but only about the correlation between the two.

Additionally, different individual characteristics are not equally prevalent amongst the respondents. While the fraction of men and women in our sample is balanced, less than $7\%$ of our respondents had heard of the scenario used in the vignettes before participating in this study. This variation in the prevalence of different individual characteristics leads to varying degrees of statistical power to detect effects of interest. In Table \ref{tbl:power_post_hoc} in the Appendix, we provide the results of a post-hoc power analysis, which details our sample's statistical power to detect medium and large sized effects. While we have sufficient power to detect large effects for all of the individual characteristics, our study is underpowered (i.e., below the standard $1-\beta=0.8$) for detecting medium-sized effects for some of the experience-related characteristics. Hence, one should not interpret the lack of a statistically significant association between an individual characteristic and fairness judgments as evidence that there is no association between the two. It is possible that associations with smaller effect sizes were not identified for the less prevalent, and consequently underpowered personal experiences.

Finally, we studied the dimensions of diversity in fairness perceptions of a sample of U.S.-based respondents for the task of making bail decisions, inspired by the COMPAS tool. We utilize the COMPAS tool as a case study since it is an example of a societally consequential machine learning algorithm that is applied in the real world. We focus on U.S.-based respondents to ensure that they have a basic understanding of the U.S. legal system in which the COMPAS tool is applied. However, prior research on concepts of diversity recognizes that the relevant dimensions of diversity may vary across contexts \cite{fazelpour2022diversity, steel2018multiple}. Hence, as a promising direction for future research, we encourage the development of a cohesive theory of human reasoning about algorithmic fairness, including the study of additional decision-making scenarios and non-U.S. populations.
\section{Results} \label{sec:results}

\begin{figure}[t]
    \centering
    \begin{subfigure}[t]{0.49\textwidth}
        \centering
        \includegraphics[width=1\columnwidth]{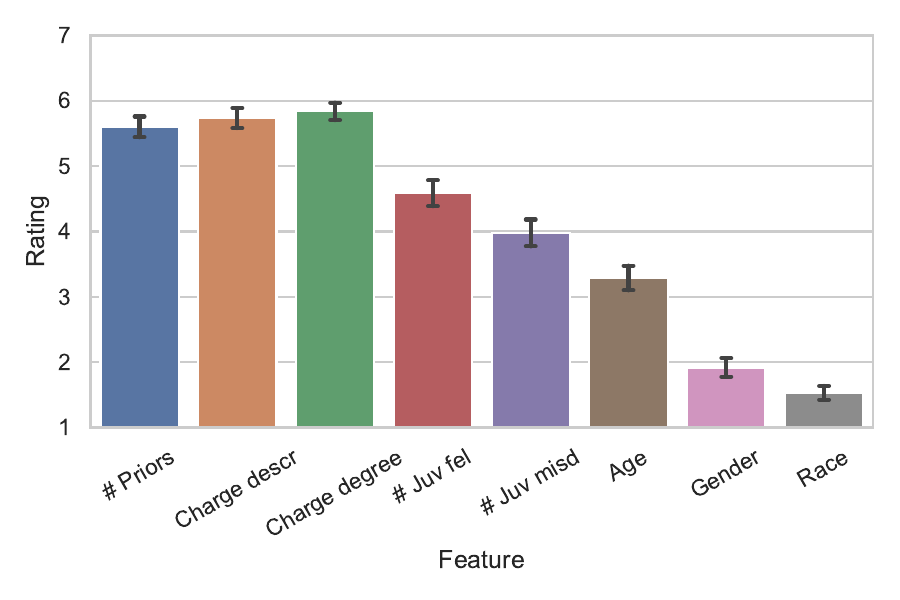}
        \caption{Mean fairness ratings of the eight ProPublica features.}
        \label{fig:feature_ratings}
    \end{subfigure}
    \hfill
    \begin{subfigure}[t]{0.49\textwidth}
        \centering
        \includegraphics[width=1\columnwidth]{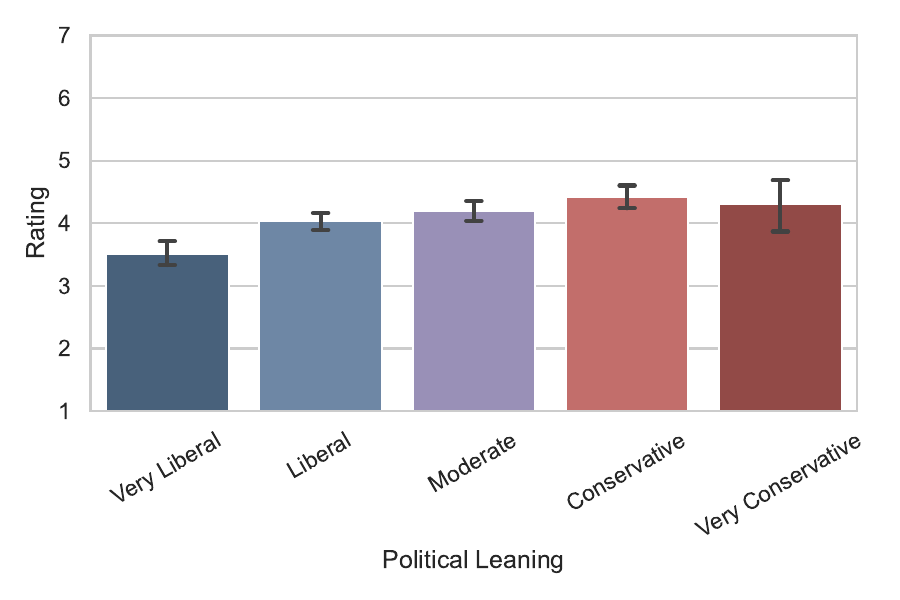}
        \caption{Mean fairness rating by respondents' political leaning.}
        \label{fig:political_leaning_across_feats}
    \end{subfigure}\\
    \begin{subfigure}[t]{1\textwidth}
        \centering
        \includegraphics[width=1\columnwidth]{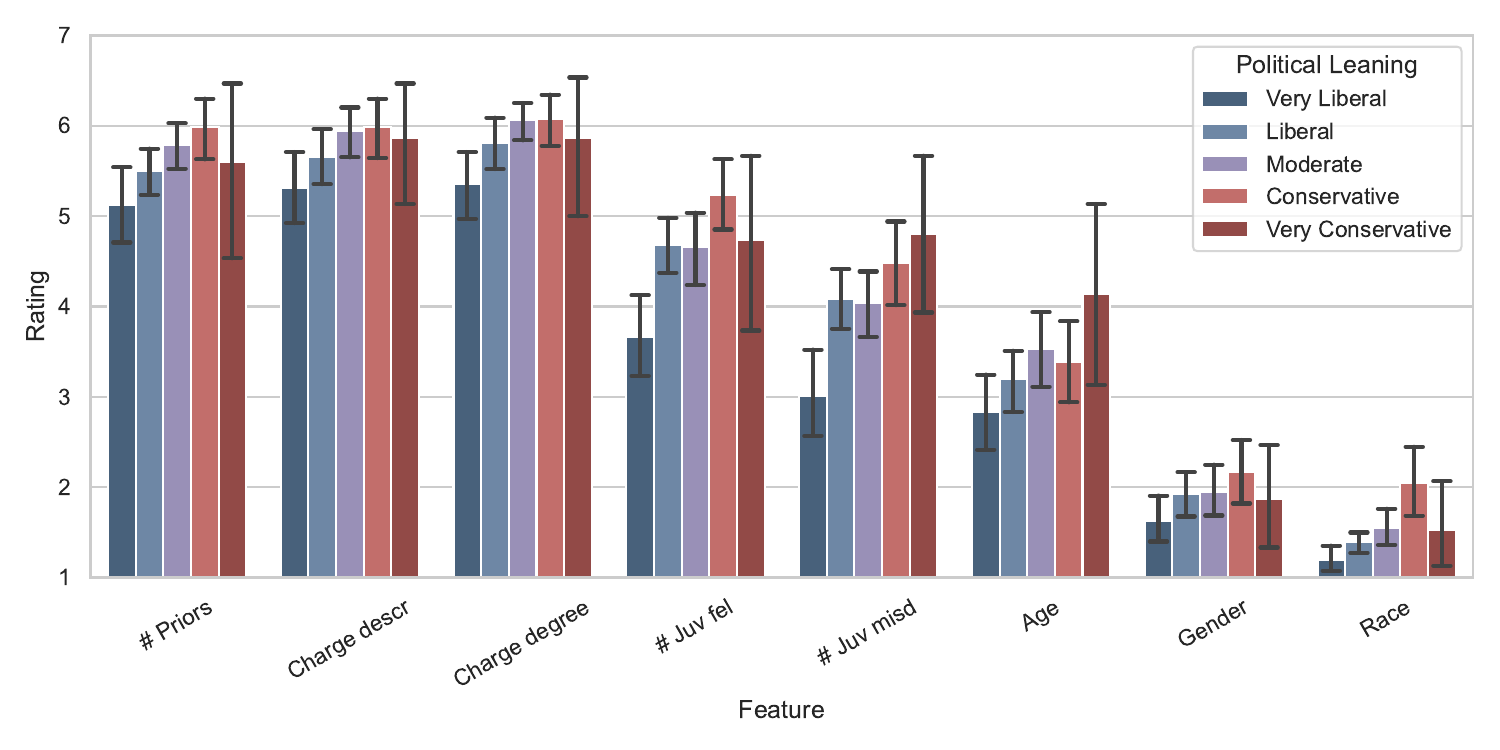}
        \caption{Mean fairness ratings of the eight features from the ProPublica dataset by respondents' political leaning.}
        \label{fig:political_leaning_per_feat}
    \end{subfigure}
    \caption{Mean fairness ratings on a 7-point Likert scale, with $95\%$ CI error bars. Larger values indicate higher perceived fairness.}
    \label{fig:all_means}
\end{figure}

When asked to determine if it is fair to use the eight ProPublica features for making bail decisions, respondents provided an average response of 4.06 out of 7---close to the midpoint of the Likert scale. While the respondents' fairness ratings averaged across features are neutral, they differ greatly between features, as shown in Figure \ref{fig:feature_ratings}. In line with the findings of \citet{grgic2018beyond}, features directly related to the bail decision (current arrest charge and adult criminal history) are considered largely fair to be used. In contrast, distantly related features (juvenile criminal history) are considered less fair, and unrelated sensitive features (age, gender, race) are perceived as unfair for making bail decisions. These descriptive observations are further corroborated by the regression in Table \ref{tbl:regression_per_feat}, where the constant terms (i.e., the estimated y-axis intercepts) vary between a minimum of 1.078 for race and a maximum of 5.673 for charge description.

\subsection{Average Pattern Across Features} \label{subsec:across_feats}

We first examine the association between respondents' individual characteristics and perceptions of fairness averaged across all eight features. We employ a mixed-effects linear regression model with fairness ratings as the dependent variable and the respondents' demographics and personal experiences as independent variables. To control for asking each respondent about all eight features (i.e., for repeated measures) we included a random effects term for respondents.

Table \ref{tbl:regression_across_feats} shows the results of this analysis. We find that political leaning is significantly correlated with fairness judgments. On average, left-leaning respondents rated using all features as less fair than right-leaning respondents. The regression model estimates that for each step on the 5-point scale from ``Very liberal'' to ``Very conservative'', the average fairness rating increases by 0.236 points, suggesting an approximate 1-point difference between the extremes. Figure \ref{fig:political_leaning_across_feats} illustrates mean perceived fairness by political leaning with the corresponding 95\% confidence intervals. The figure shows a clear positive correlation up until very conservative respondents, who are rare in our sample and hence have a much larger standard error. No other individual characteristic was found to be significantly correlated with fairness judgments in this analysis. 

We further explored if the effects of other individual attributes may be subsumed by the effect of political leaning. To investigate this, we trained another model where we removed political leaning from the set of independent variables. We found that in this model the respondents' age exhibits a significant positive association with fairness ratings (coef $=0.086$, $p=.045$), while having attended a bail hearing exhibits a borderline significant negative association (coef $=-0.3$, $p=.051$), as shown in Table \ref{tbl:regression_across_feats_no_pl} in the Appendix.

\xhdr{Summary} When considering the participants' responses averaged across features, we find support for Hypothesis 1 (political leaning), but we find no support for Hypothesis 2 (gender).

\subsection{Feature-Specific Patterns} \label{subsec:per_feat}
To analyze the association between respondents' characteristics and their judgments of specific features, we employ a multivariate linear model shown in Table \ref{tbl:regression_per_feat}. Unlike the model in Section \ref{subsec:across_feats}, which uses a single dependent variable and does not model the variation across features, this model models the eight dependent variables---the fairness ratings of the eight features---simultaneously. Again, we use the respondents' individual characteristics as independent variables. 

In line with our findings from the previous subsection, political leaning is positively correlated with the perceived fairness of most features. Figure \ref{fig:political_leaning_per_feat} presents this association for each feature separately. We observe that the effect size varies significantly across features. As shown in Table \ref{tbl:regression_per_feat}, the effect size is the largest and the p-values are the smallest ($p<.001$) for features related to juvenile crimes. For each step on the political spectrum (from ``Very liberal'' to ``Very conservative''), fairness ratings increase by 0.353 and 0.405 points for juvenile felonies and misdemeanors respectively. This corresponds to an estimated difference of approximately 1.5 points between very conservative and very liberal respondents. For the remaining features with a significant association, this effect is smaller, with a difference of approximately 0.8 points between respondents on different ends of the political spectrum.
Gender is the only feature not significantly associated with political leaning. 

Additionally, we find that having attended a bail hearing is negatively correlated ($p<.01$) with the perceived fairness of using features related to a defendant's juvenile crimes. Respondents who have attended a bail hearing rate the fairness of these features more than two-thirds of a point lower than those who have not.
Finally, respondents who identify as men rated using information about a defendant's race by a quarter of a point fairer than others ($p<.05$).

Again, we explored if the respondents' political leaning may be subsuming the effects of their other individual characteristics, by training a model where political leaning is excluded from the set of independent variables. We found that in this model the respondents' age exhibits a significant positive association with the perceived fairness of using information about a defendant's juvenile felonies (coef $=0.195$, $p=.015$), as shown in Table \ref{tbl:regression_per_feat_no_pl} in the Appendix.

\begin{landscape}
\mbox{}
\vfill
\begin{table}[h]
\centering
\def\sym#1{\ifmmode^{#1}\else\(^{#1}\)\fi}
\begin{tabulary}{2\linewidth}{L|R|R|R|R|R|R|R|R}
\toprule
& {\bf \# Priors} & {\bf Charge desc.} & {\bf Charge deg.} & {\bf \# Juv fel} & {\bf \# Juv misd} & {\bf Age} & {\bf Gender} & {\bf Race} \\
\hline
Age & 0.0624 & 0.00239 & -0.0402 & 0.140 & 0.0337 & 0.106 & 0.0797 & 0.0103 \\
 & (0.0650) & (0.0666) & (0.0603) & (0.0791) & (0.0823) & (0.0823) & (0.0581) & (0.0443) \\
Political leaning & \textbf{0.207}\sym{**} & \textbf{0.206}\sym{**} & \textbf{0.206}\sym{**} & \textbf{0.353}\sym{***}& \textbf{0.405}\sym{***}& \textbf{0.216}\sym{*} & 0.104 & \textbf{0.192}\sym{***}\\
 & (0.0706) & (0.0723) & (0.0655) & (0.0858) & (0.0894) & (0.0894) & (0.0631) & (0.0482) \\
Bachelor's or above & -0.163 & 0.0748 & -0.0740 & 0.00458 & 0.140 & 0.180 & 0.127 & -0.0514 \\
 & (0.167) & (0.171) & (0.155) & (0.203) & (0.211) & (0.211) & (0.149) & (0.114) \\
Male & -0.00124 & -0.298 & 0.00316 & 0.180 & 0.0986 & 0.0627 & 0.0821 & \textbf{0.267}\sym{*} \\
 & (0.160) & (0.164) & (0.148) & (0.194) & (0.203) & (0.202) & (0.143) & (0.109) \\
White & -0.0596 & -0.214 & -0.110 & -0.412 & -0.231 & 0.0530 & 0.169 & 0.0533 \\
 & (0.181) & (0.185) & (0.168) & (0.220) & (0.229) & (0.229) & (0.162) & (0.123) \\
\hline
Heard of scenario & 0.0939 & -0.278 & 0.254 & 0.246 & 0.0289 & -0.288 & -0.159 & -0.148 \\
 & (0.322) & (0.330) & (0.299) & (0.392) & (0.409) & (0.408) & (0.288) & (0.220) \\
Legal profession - you & 0.00536 & 0.126 & -0.279 & -0.178 & -0.00699 & -0.0242 & -0.166 & -0.203 \\
 & (0.319) & (0.327) & (0.296) & (0.388) & (0.405) & (0.404) & (0.286) & (0.218) \\
Legal profession - fr \& rel & -0.0499 & -0.0363 & 0.0422 & -0.251 & -0.0347 & 0.0217 & -0.0649 & 0.258 \\
 & (0.196) & (0.201) & (0.182) & (0.238) & (0.248) & (0.248) & (0.175) & (0.134) \\
Attended bail hearing & -0.317 & 0.0278 & -0.110 & \textbf{-0.824}\sym{**} & \textbf{-0.700}\sym{*} & 0.0506 & -0.0485 & -0.131 \\
 & (0.231) & (0.236) & (0.214) & (0.280) & (0.292) & (0.292) & (0.206) & (0.157) \\
Served on jury & -0.0288 & -0.0115 & 0.199 & 0.153 & 0.150 & -0.000934 & -0.0251 & -0.0906 \\
 & (0.198) & (0.203) & (0.183) & (0.241) & (0.251) & (0.250) & (0.177) & (0.135) \\
\hline
Constant & \textbf{5.367}\sym{***}& \textbf{5.673}\sym{***}& \textbf{5.665}\sym{***}& \textbf{4.149}\sym{***}& \textbf{3.386}\sym{***}& \textbf{2.597}\sym{***}& \textbf{1.431}\sym{***}& \textbf{1.078}\sym{***}\\
 & (0.249) & (0.255) & (0.231) & (0.303) & (0.315) & (0.315) & (0.223) & (0.170) \\
\bottomrule
\end{tabulary}
\caption{Multivariate linear (structural) model. Dependent variables (columns): fairness ratings of the eight features on a 7-point Likert scale. Larger values indicate higher perceived fairness. Independent variables (rows): respondents' \textbf{demographics and experiences}. Reference groups for the ordinal variables age and political leaning are ``18-24'' and ``very liberal'' respectively. Number of observations per dependent variable = 329. Standard errors in parentheses. *** p < .001, ** p < .01, * p < .05.}
\label{tbl:regression_per_feat}
\end{table}
\vfill
\end{landscape}

Finally, it is worth noting that while individuals with different political views differ systematically in their absolute fairness assessments, the order in which they rank algorithmic features---from most fair to use to least fair---appears consistent across different political groups. Figure \ref{fig:ranking_heatmap} in the Appendix shows the similarity between the rankings by respondents with different political leanings. The rankings are derived from the features' mean fairness ratings by political leaning, and the similarity is quantified using Kendall's Tau ($\tau$). We observe that all pairs of political leanings exhibit a high correlation in their rankings of features, with Kendall's $\tau$ values close or equal to 1. That is, across the political spectrum, respondents perceive using information about the defendant's current charge as more fair than using information about the their juvenile criminal history.

\xhdr{Summary} We find support for Hypothesis 1 (political leaning), and weak partial support for Hypothesis 2 (gender). We find no support for Hypotheses 3a) (ego- and group- justifying behavior) and 3b) (system- justifying behavior).
\section{Discussion} \label{sec:discussion}
Here, we discuss our key findings related to the dimensions of diversity explored in this work – political views, demographics, and personal experiences – as well as the implications of these findings.

\xhdr{Political Views}
Consistent with prior findings in Moral Foundations Theory \cite{graham2013moral,efferson2017influence} and research on diversity in perceptions of algorithmic fairness \cite{grgic2018beyond}, the respondents' political views were found to be significantly correlated with their fairness judgments for most features. The more conservative an individual is, the more fair they perceive using most features for bail decisions. This trend of conservatives to view information about individuals as fair to use in making decisions about them also aligns with the framework of individualist vs. structuralist beliefs, which has been primarily explored in studies of racism, poverty, and crime in the U.S. \cite{hunt2004race, zucker1993conservatism, carroll1987sentencing}. Conservatives tend to believe in ``individualist'' explanations for outcomes---which emphasize individual responsibility---as compared to liberals, who tend to make structuralist attributions, emphasizing how social structures create outcomes. 

It is worth noting that the magnitude and statistical significance of this association varies across features. The effect is the largest for features related to a defendant's juvenile criminal history. This observation may partially be explained by the deservingness heuristic. People's welfare allocation preferences are found to vary depending not only on their political values, but also on the welfare recipient's perceived deservingness. When information about a recipients dersevingness of welfare is available, it is found to outweigh the impact of political values \cite{petersen2011deservingness}. Some of the features we consider---such as race and gender---are closely related to perceptions of deservingness in welfare allocation settings \cite{jilke2018clients}. Other features---such as information about the current charge degree and adult criminal history---may be closely related to perceptions of deservingness in the task at hand, since they are perceived as the most relevant, reliable and fair features to be used in the bail decision-making setting \cite{grgic2018human}. This may explain why the perceived fairness of these features exhibits a smaller association (or, for gender, no association) with the respondents' political leaning.

\xhdr{Demographic Factors}
We found no evidence of demographic factors having a consistent significant association with algorithmic fairness judgments. Our finding is in line with the work on perceptions of algorithmic fairness by \citet{araujo2020ai} and \citet{wang2020factors}, who also found little effect of demographic factors. 

The significance of political views and the lack of support for other demographic factors is in line with reports from the Pew Research Center, who find the same patterns in the context of predictors of political attitudes in the U.S \cite{pew2019polarization}. Recent work by \citet{iyengar2019origins} on affective polarization argues that ``[a]s partisan and ideological identities became increasingly aligned, other salient social identities, including race and religion, also converged with partisanships''. Hence political leaning may in fact be subsuming other sociodemographic dimensions. Some of our exploratory analyses hint that this may be the case. We found that when we do not control for the respondents' political leaning, their age exhibits some association with fairness ratings. This pattern may be explained in part by the correlation between the respondents' age and political leaning in our sample.

Building upon the social science literature, we also hypothesized that respondents' demographics may relate to the perceived fairness of using those same demographics---age, gender and race---in the decision-making task. We hypothesized that the direction of this effect may be egocentric, in line with research on egocentric interpretations of fairness~\cite{thompson1992egocentric}, or the opposite, in line with System Justification Theory \cite{jost2019quarter}. Our results do not offer support for either of these hypotheses. These findings are contrary to those of \citet{pierson2017gender}, who found an ego- and group-justifying association for gender, albeit in a different setting: algorithms for recommending college courses. This may suggest that such associations are scenario-dependent, and future work may seek to explore the relationship between AI fairness beliefs and sociodemographics in other contexts.

Nonetheless, we identified one weakly significant association between demographics and the perceived fairness of using certain features. Namely, we found that men perceive using information about race to be more fair than women do. This finding is in line with prior research on Moral Foundations Theory, which found that women are more concerned about harm and fairness issues than men, even when controlling for their political views \cite{graham2011mapping}. Given that much of modern-day discourse on fairness in policing concerns racial discrimination, especially in the context of algorithms utilized in the law enforcement domain \cite{propublica_story, richardson2019dirty}, it is possible that race is the most salient fairness issue in the context of our study. Our finding may also be connected to prior research that found that women may exhibit less racial bias than men in certain policy questions \cite{hughes2003gender}. However, these associations were found to be small and inconsistent across studies, perhaps explaining the weakly significant effect we identified.

\xhdr{Personal Experiences}
Only one personal experience was found to be significantly associated with fairness judgments: having attended a bail hearing was correlated with a lower likelihood of rating certain features as fair for making bail decisions. Prior research on the effect of past experiences on fairness preferences predominantly focused on the effects of negative and traumatic experiences, such as economic hardships \cite{margalit2019political,giuliano2014growing} or natural disasters \cite{gualtieri2018natural}. Having attended a bail hearing is arguably the only of the four experiences that can be classified as negative or traumatic, which may explain why it is the only feature for which we identified an association with fairness judgments. 

Additionally, out of the four prior experiences we considered, having attended a bail hearing is the most closely related to the bail decision-making scenario. This suggests that personal experiences may need to be directly tied to the decision at hand in order to lead to an effect. Depending on the respondent's role in the bail hearing \footnote{The question ``Have you ever attended a bail hearing?'' did not specify the respondent's role in the hearing (e.g., victim, defendant, attorney, ...), in order to minimize the sensitivity and intrusiveness of our experiment.} this may be an example of an egocentric effect~\cite{thompson1992egocentric} or ego- and group-justifying behavior \cite{jost2019quarter}.

Finally, when attending a bail hearing, respondents may have acquired additional information that led them to believe that using information about juvenile criminal history is unfair in such settings, e.g., by learning that juvenile court records have a confidential status in many settings \cite{nelson1997release}. This raises the question why this pattern was not observed for other legally protected attributes, such as race. A potential explanation could be that the protected status of race is already more broadly applied and widely known than that of juvenile crimes, hence diminishing the learning effect.

\xhdr{Implications}
The judgment of which features are fair to use in a given decision-making task is a moral decision, requiring human background knowledge and societal context that may often not be present in the data provided. The appropriate person(s) to make this judgment may vary by setting: policy makers, domain experts, algorithm designers, oversight boards, or even the general population (a descriptive ethics approach~\cite{sep-morality-definition}). 

With increasing calls for diversity in the design, oversight and governance of algorithms, it is important to understand which dimensions of diversity are associated with significant variance across judgments in the context of interest, in order to ensure adequate representation of diverse views. Our results indicate that, at least in the U.S., political leaning may be a critical factor of diversity to consider. This may be timely given observations of increasingly politically homogeneous media outlets \cite{prior2013media}, corporate boards \cite{fos2021political, hoang2020polarized}, and academic institutions \cite{langbert2018homogenous}.

Our findings also highlight the importance of considering dimensions of diversity that go beyond demographic and ideological differences. Our work and prior research in human-computer interaction and social psychology all suggest that closely-related personal experiences (e.g., attending a bail hearing) may influence people's moral judgments.

Specifically, in our work, we identified the experience of having attended a bail hearing as one that significantly impacts fairness judgments in the context of machine-assisted bail decisions. However, the relevant prior experiences will likely vary across settings. Further, it is not yet clear when such personal experiences offer benefits---access to otherwise unknown knowledge or context---and when they may lead to a (negative) bias, as in prior work where crowdworkers rated AI that rated them negatively as less fair, even if the decisions were correct~\cite{wang2020factors}. Thus, further work is needed to better understand in what contexts ensuring diversity with respect to lived experiences is desirable, how closely related personal experiences must be to a given decision context to be relevant, and how the uniqueness of a decision context may influence the relevance of such experiences. We invite future work that moves toward building a cohesive theory that explains the effects of lived experiences on fairness judgments.

\xhdr{Conclusion}
The design of ML algorithms aligned with societal moral values is inherently an interdisciplinary endeavor. This requires technical contributions from the ML community, guidelines from policy makers, and research in the social sciences, HCI and related fields that can help gather and understand insights about perceptions of algorithmic fairness. In this work, we identify a set of sociodemographic and experience factors that are associated with people's judgments about the fairness of using various features for making bail decisions. However, relevant dimensions of diversity may differ in other decision-making settings and cultural contexts. We hope this paper will inspire future research about the relevant dimensions of diversity in the governance of AI systems, to help ensure that the diverse perspectives of all stakeholders can be appropriately represented in discussions about societally consequential algorithms.

\begin{acks}
  This research was supported in part by a European Research Council (ERC) Advanced Grant for the project "Foundations for Fair Social Computing", funded under the European Union’s Horizon 2020 Framework Programme (grant agreement no. 789373). Adrian Weller acknowledges support from a Turing AI Fellowship under grant EP/V025279/1, The Alan Turing Institute, and the Leverhulme Trust via CFI.
\end{acks}

\bibliographystyle{ACM-Reference-Format}
\bibliography{egofair}

\appendix
\clearpage
\pagebreak
\clearpage
\section{Appendix} \label{sec:appendix}

\begin{figure}[h]
    \centering
    \includegraphics[width=0.65\columnwidth]{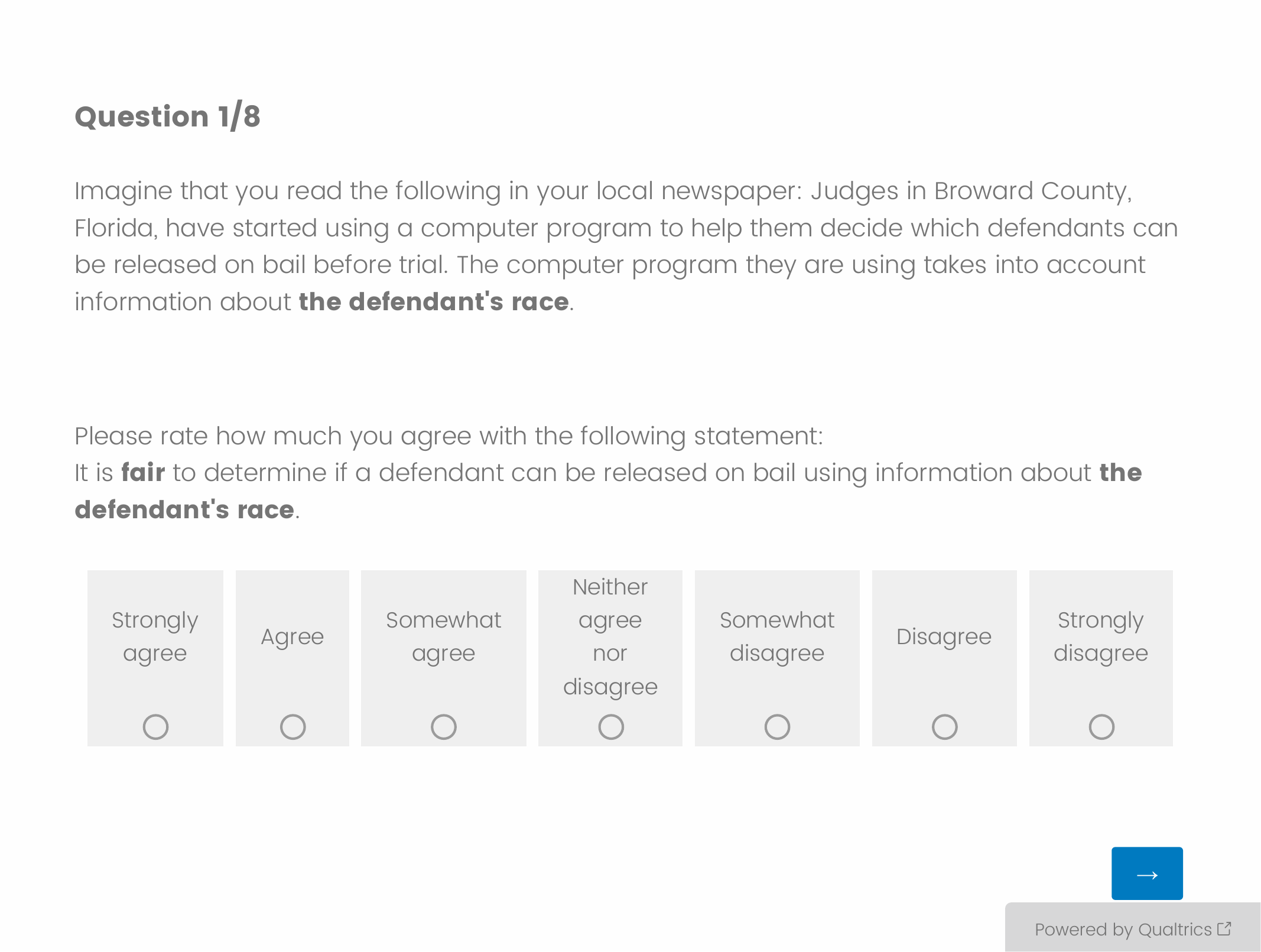}
    \caption{Survey screenshot: question and response options for the feature \emph{race}.}
    \label{fig:question}
\end{figure}

\begin{figure}[h]
    \centering
    \includegraphics[width=0.65\columnwidth]{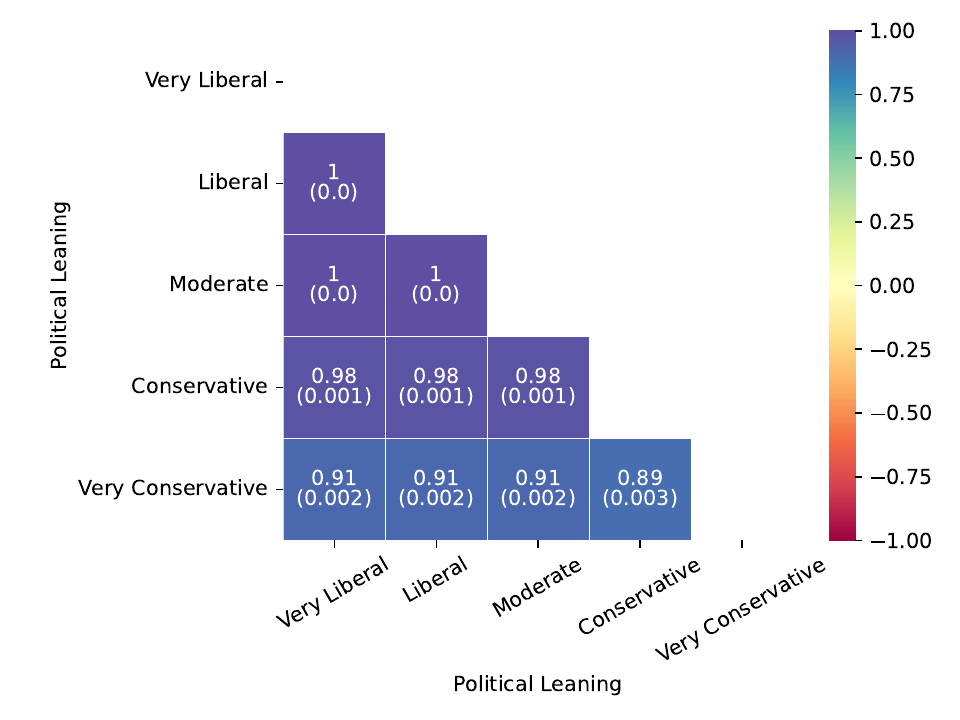}
    \caption{Similarity between the rankings of features by respondents with different political views. Features were ranked based on their mean fairness ratings. The heatmap shows the Kendall's Tau ($\tau$) of the rankings and corresponding p-values in parentheses. $\tau$ values close to 1, 0 and -1 indicate positive, no, and negative correlation, respectively.}
    \label{fig:ranking_heatmap}
\end{figure}

\clearpage

\begin{table}[t]
\centering
\small
\begin{tabulary}{\textwidth}{p{.11\textwidth}|p{.55\textwidth}|p{.26\textwidth}}
    \toprule
    \textbf{Variable Type} & \textbf{Question} & \textbf{Response Options} \\
    \hline
    \textbf{Fairness judgments} 
    & Imagine that you read the following in your local newspaper: Judges in Broward County, Florida, have started using a computer program to help them decide which defendants can be released on bail before trial. The computer program they are using takes into account information about \textbf{<feature>}. 
    & 7-point Likert scale, from ``Strongly agree'' to ``Strongly disagree'' \\
    & Please rate how much you agree with the following statement: It is \textbf{fair} to determine if a defendant can be released on bail using information about \textbf{<feature>}. & \\
    \hline
    \textbf{Demographics}
    & Which of the following best identifies your gender? &
    Standard response options, and \\
    & What is your age? & ``Other'' and ``Prefer not to respond'' \\
    & What is the highest degree or level of school you have completed? If currently enrolled, highest degree received. & \\
    & Which of the following best defines your political view? & \\
    & Which of the following best identifies your race/ethnicity? & \\
    \hline
    \textbf{Personal experiences} 
    & Before taking this survey, have you ever heard or read anything related to the scenario described in this survey? & ``Yes'', ``No'', ``Prefer not to respond'' \\
    & Do you hold a job or have education in law, law enforcement, criminology, or a related field?  & \\
    & Do you have any close friends or relatives who work or have an education in law, law enforcement, criminology, or a related field? & \\
    & Have you ever attended a bail hearing? & \\
    & Have you ever served on a jury? & \\
    \bottomrule
    \end{tabulary}
    \caption{Survey questions and response options for the dependent (fairness judgments) and independent (demographics and personal experiences) variables used in this study.}
    \label{tbl:survey_questions}
\end{table}

\clearpage


\begin{table}[t]
    \centering
    \begin{tabular}{l|r|r}
    \toprule
    & \multicolumn{2}{c}{\bf Effect size} \\
    \hline
    {\bf Individual} & {\bf Medium} & {\bf Large} \\
    {\bf Characteristic} & ($d=0.5$) & ($d=0.8$)\\
    \hline
    Male vs Non-Male & 0.992 & 0.999 \\
    White vs POC & 0.974 & 0.999 \\
    $\geq$ BA vs $<$ BA & 0.990 & 0.999 \\
    \hline
    Heard of scenario & 0.598 & 0.942 \\
    Legal profession -- you & 0.648 & 0.963 \\
    Legal profession -- fr \& rel & 0.961 & 0.999 \\
    Attended bail hearing & 0.877 & 0.998 \\
    Served on jury & 0.958 & 0.999 \\
    \bottomrule
    \end{tabular}
    \caption{Post-hoc power analysis of binary variables. The table shows the statistical power ($1-\beta$) to detect medium (Cohen's $d=0.5$) and large (Cohen's $d=0.8$) effects, given the prevalence of different individual characteristics in our sample, using a two-tailed Wilcoxon-Mann-Whitney test with $\alpha=0.05$. Values closer to 1 represent a lower probability of making a type II error (false negative), with $0.8$ being the standard cutoff.}
    \label{tbl:power_post_hoc}
\end{table}

\begin{table}[h]
\centering
\def\sym#1{\ifmmode^{#1}\else\(^{#1}\)\fi}
\begin{tabulary}{\linewidth}{L|R|R}
\toprule
& {\bf Coef.} & {\bf SE} \\
\hline
Age & \textbf{0.0860}\sym{*} & (0.0429)\\
Bachelor's or above & -0.0711 & (0.110)\\
Male & 0.0990 & (0.107)\\
White & -0.0742 & (0.121)\\
\hline
Heard of scenario & -0.151 & (0.215)\\
Legal profession - you & -0.0259 & (0.213)\\
Legal profession - fr \& rel & -0.0181 & (0.131)\\
Attended bail hearing & -0.301 & (0.154)\\
Served on jury & -0.0115 & (0.132)\\
\hline
Constant & \textbf{4.028}\sym{***}& (0.150)\\
\bottomrule
\end{tabulary}
\caption{Linear mixed model with a random effects term for respondents. Dependent variable: fairness ratings on a 7-point Likert scale with larger values indicating higher perceived fairness. Independent variables (rows): respondents' \textbf{demographics and experiences, excluding political leaning}. The reference group for the ordinal variable age is ``18-24''. Number of observations = 2632. Standard errors in parentheses. *** p < .001, ** p < .01, * p < .05.}
\label{tbl:regression_across_feats_no_pl}
\end{table}

\clearpage
\begin{landscape}
\mbox{}
\vfill
\begin{table}[h]
\centering
\def\sym#1{\ifmmode^{#1}\else\(^{#1}\)\fi}
\begin{tabulary}{2\linewidth}{L|R|R|R|R|R|R|R|R}
\toprule
& {\bf \# Priors} & {\bf Charge desc.} & {\bf Charge deg.} & {\bf \# Juv fel} & {\bf \# Juv misd} & {\bf Age} & {\bf Gender} & {\bf Race} \\
\hline
Age & 0.0944 & 0.0343 & -0.00829 & \textbf{0.195}\sym{*} & 0.0965 & 0.140 & 0.0958 & 0.0401 \\
 & (0.0649) & (0.0664) & (0.0603) & (0.0799) & (0.0836) & (0.0818) & (0.0575) & (0.0448) \\
Bachelor's or above & -0.251 & -0.0131 & -0.162 & -0.146 & -0.0324 & 0.0878 & 0.0821 & -0.133 \\
 & (0.166) & (0.170) & (0.154) & (0.204) & (0.214) & (0.209) & (0.147) & (0.114) \\
Male & 0.0423 & -0.255 & 0.0465 & 0.254 & 0.184 & 0.108 & 0.104 & \textbf{0.308}\sym{**} \\
 & (0.161) & (0.165) & (0.150) & (0.199) & (0.208) & (0.203) & (0.143) & (0.111) \\
White & -0.0423 & -0.197 & -0.0930 & -0.383 & -0.197 & 0.0711 & 0.178 & 0.0694 \\
 & (0.183) & (0.188) & (0.170) & (0.226) & (0.236) & (0.231) & (0.162) & (0.126) \\
\hline
Heard of scenario & -0.0107 & -0.382 & 0.150 & 0.0673 & -0.176 & -0.397 & -0.212 & -0.245 \\
 & (0.325) & (0.332) & (0.302) & (0.400) & (0.418) & (0.409) & (0.288) & (0.224) \\
Legal profession - you & 0.0621 & 0.183 & -0.223 & -0.0814 & 0.104 & 0.0350 & -0.137 & -0.150 \\
 & (0.323) & (0.331) & (0.300) & (0.398) & (0.416) & (0.407) & (0.286) & (0.223) \\
Legal profession - fr \& rel & -0.0532 & -0.0396 & 0.0388 & -0.256 & -0.0413 & 0.0182 & -0.0666 & 0.255 \\
 & (0.198) & (0.203) & (0.184) & (0.244) & (0.256) & (0.250) & (0.176) & (0.137) \\
Attended bail hearing & -0.355 & -0.0106 & -0.148 & \textbf{-0.890}\sym{**} & \textbf{-0.776}\sym{**} & 0.0103 & -0.0680 & -0.167 \\
 & (0.233) & (0.239) & (0.217) & (0.287) & (0.301) & (0.294) & (0.207) & (0.161) \\
Served on jury & -0.0767 & -0.0592 & 0.152 & 0.0715 & 0.0557 & -0.0509 & -0.0492 & -0.135 \\
 & (0.200) & (0.204) & (0.186) & (0.246) & (0.257) & (0.252) & (0.177) & (0.138) \\
\hline
Constant & \textbf{5.682}\sym{***}& \textbf{5.986}\sym{***}& \textbf{5.979}\sym{***}& \textbf{4.688}\sym{***}& \textbf{4.003}\sym{***}& \textbf{2.926}\sym{***}& \textbf{1.589}\sym{***}& \textbf{1.371}\sym{***}\\
 & (0.227) & (0.233) & (0.212) & (0.280) & (0.293) & (0.287) & (0.202) & (0.157) \\
\bottomrule
\end{tabulary}
\caption{Multivariate linear (structural) model. Dependent variables (columns): fairness ratings of the eight features on a 7-point Likert scale. Larger values indicate higher perceived fairness. Independent variables (rows): respondents' \textbf{demographics and experiences, excluding political leaning}. The reference group for the ordinal variable age is ``18-24''. Number of observations per dependent variable = 329. Standard errors in parentheses. *** p < .001, ** p < .01, * p < .05.}
\label{tbl:regression_per_feat_no_pl}
\end{table}
\vfill
\end{landscape}

\end{document}